\documentstyle[12pt]{article}
\catcode`\"=\active \let"=\" \let\3=\ss 
\renewcommand{\baselinestretch}{1.2}
\begin{document}
\title{Influence of spin on the persistent current of strongly
interacting electrons}
\author{Wolfgang H\"ausler\\
{\it I.~Institut f"ur Theoretische Physik, Universit"at Hamburg}\\
{\it Jungiusstr. 9, 20355 Hamburg, F.~R.~G.}}
\date{Received~:\rule{3cm}{0cm}}
\maketitle
\begin{abstract}
The lowest eigenenergies of few, strongly interacting electrons in
a one--dimensional ring are studied in presence of an impurity barrier.
The persistent current $\:I\:$, periodic in an Aharonov--Bohm
flux penetrating the ring, is strongly influenced by the
electron spin. The impurity does not remove discontinuities in
$\:I\:$ at zero temperature. The total electron spin of the ground
state oscillates with the flux. Strong electron--electron interaction
enhances $\:I\:$, albeit not up to the value of the clean ring which
itself is smaller than $\:I\:$ for free electrons. $\:I\:$ disappears
on a temperature scale that depends exponentially on the electron
density. In the limit of very strong interaction the response to
small fluxes is diamagnetic.
\end{abstract}
\vspace{7mm}

\noindent
\begin{tabular}{l@{\hspace{5mm}}l}
05.30.Fk & Fermion Systems\\
71.27.+a & Strongly Correlated Electron Systems\\
73.20.Dx & Electron States in Low--dimensional Systems\\
75.30.Et & Exchange and Superexchange Interactions
\end{tabular}
\newpage
\newcommand{\bra}[1]{\raisebox{-0.3ex}{\mbox{\large\tt <}}#1
 \hspace{0.5ex}\rule[-0.5ex]{0.1ex}{0.9em}\hspace{0.5ex}}
\newcommand{\ket}[1]{\hspace{0.5ex}\rule[-0.5ex]{0.1ex}{0.9em}
 \hspace{0.5ex}#1\raisebox{-0.3ex}{\mbox{\large\tt >}}}
\newcommand{\braket}[2]{\raisebox{-0.3ex}{\mbox{\large\tt <}}#1
 \hspace{0.5ex}\rule[-0.5ex]{0.1ex}{0.9em}\hspace{0.5ex}
 #2\raisebox{-0.3ex}{\mbox{\large\tt >}}}
\renewcommand{\d}{\displaystyle}
\newcommand{\numb}{\begin{equation}\hspace{-1cm}
 \begin{array}{rcl}}\newcommand{\nume}[1]{\end{array}\label{#1}\end{equation}}
\newcommand{\Kr}[1]{\left( #1\right)}
\newcommand{\Kg}[1]{\left\{ #1\right\}}
\newcommand{\lapprox}{\mbox{\raisebox{-4pt}{$\,\buildrel<\over\sim\,$}}}
\newcommand{\gapprox}{\mbox{\raisebox{-4pt}{$\,\buildrel>\over\sim\,$}}}
\newcommand{\ve}{\varepsilon}
\renewcommand{\O}{\mbox{$\cal O$}}
\newfont{\indx}{cmbx8}
\section{Introduction}
The appearance of persistent currents is a consequence of the coherent
electron motion in a ring \cite{percurr}. In contrast to transport
currents this is an equilibrium property also of non--superconducting
electrons in small rings enclosing an Aharonov--Bohm flux. The
persistent current is observable at sufficiently low
temperatures \cite{mailly93} even in the presence of disorder
\cite{levi,chandrasekar}. Only sophisticated SQUID--techniques
allow to separate the equilibrium magnetization from the
externally applied field. Much larger magnitudes were found for
the currents $\:I\:$ than theoretically predicted \cite{oppen91,altshuler91}.
Non--interacting electrons can explain neither the magnetization
observed in $10^7$ rings \cite{levi} nor in single rings
\cite{chandrasekar}.

Contributions from electron--electron interaction, estimated on
the Hartree--Fock level \cite{eckern93}, improve the single
electron estimates though the results are not yet conceived as
being completely satisfying. Additionally, at reduced dimensionalities
or electron densities, as it appears in semiconducting
rings \cite{mailly93}, a mean field description becomes
unreliable \cite{ando82,hausler93}.

The persistent current of interacting, spinless electrons on a
one--dimensional (1D) clean ring has been studied within a
Luttinger liquid model \cite{loss92}. The question in how far
interaction modifies the magnitude of the current in a clean ring
has been discussed controversely \cite{loss92,weidenmueller93}.
The sign of the magnetic response is found to be the same as
for spinless non--interacting electrons.

The influence of strong Coulomb interaction in a disordered ring
has not yet been clarified. In a continuous ring the interaction
is predicted to enhance the current \cite{weidenmueller93}, while
reduced currents, even below their value in the absence of
interaction, were found from a discrete Hubbard type model
where finite ranges of the interaction have been considered
\cite{montambaux93c}.

A new, interesting question in the presence of interaction is in how
far the electron spin is important. This has been investigated up to
now only for a clean Hubbard model \cite{fowler92a}. Spinless impurity
scatterers and a constant vector potential conserve the total electron
spin. Energy levels with different spins are not mixed by the impurities
and can intersect when the flux changes. Only energy levels of the same
spin repel each other. It is not clear in how far random matrix theory
can be applied even within the blocks of given total spins
\cite{berkovits94}. A generalization of the concept of the
Thouless energy \cite{imry91} is presumably needed
in presence of electron--electron interaction.

In the present work interacting electrons with spin on a 1D,
continuous ring consider that contains an impurity barrier. Strong
correlations, found at moderately low electron densities, can be
included by the present approach. Like in the context of quantum
dots \cite{hausler93,pocket} classical, Wigner crystallized electrons
are taken as starting point for which quantum corrections are
calculated using group theoretical means.

The sign of the susceptibility to small fluxes will shown to be
not only dependent on the parity of the electron number but also
on the strengths of impurity and interaction. The total spin
values in the ground state can periodically vary with flux. At
very strong interaction the response turns out to be always
diamagnetic. Backscattering from the impurity reduces the
probability for the electrons to circulate around the ring which
reduces the magnitude of the current. The electron spin can
cause the persistent current to increase with increasing
interaction strength. The interaction changes the energy level
spectrum considerably and therefore the temperature scale for the
persistent current to disappear compared to non--interacting or
spinless electrons.
\section{Model}\label{model}
The model describing $\:N\:$ interacting electrons on a
quasi one--dimensional ring penetrated by an Aharonov--Bohm flux
$\:\phi\;\frac{h}{e}\:$ ($\:\phi\:$ is the number of flux
quanta) is in polar coordinates,
\begin{equation}\label{1dmodel}
H=B\sum_{j=1}^{N}(-i\frac{\partial}{\partial\vartheta_j}-
\phi)^2+\frac{1}{2}\sum_{j,j'}
w(|\vartheta_j-\vartheta_{j'}|)+\sum_j v(\vartheta_j)\quad,
\end{equation}
where $\:B=h^2/2mL^2\:$ is the rotational constant of a mass $\:m\:$ on
the ring of circumference $\:L\:$. For simplicity only one, repulsive
impurity barrier, $\:v(\vartheta)\:$, is assumed to be present with
a range shorter than the mean electron separation $\:a=L/N\:$.
The range of the repulsive interaction $\:w(|\vartheta|)\ge 0\:$
is assumed to be larger than the width of the ring. An example would
be the Coulomb interaction
\begin{equation}\label{coulomb}
w(|\vartheta|)=2\pi e^2/\ve L\sqrt{\vartheta^2+b^2}
\end{equation}
in a ring of transversal width $\:b\,L/2\pi\ll L\:$.
The curvature of the ring can be neglected if $\:N\gg 2\pi\:$.

For low densities, $\:a\gapprox a_{\rm B}=\ve\hbar^2/ me^2\:$
($\:\ve\:$ is the dielectric constant), the electrons form a
`Wigner molecule', due to the rapid decay of the kinetic energy
compared to the repulsion (\ref{coulomb}). The impurity prohibits
free overall rotation and well defined electron sites on the ring
become energetically favourable. At very large $\:a\:$ the ground
state is independent of the individual spin orientations
$\:2^N$--fold degenerate. Increasing densities remove this
degeneracy. The ensuing energy splittings $\:\Delta\:$ are due to
tunneling and proportional to the rates for the classically
forbidden transitions of electrons exchanging positions. The
resulting low energy excitation spectrum can be obtained within
the pocket state approximation (PSA) \cite{pocket}.
\section{Pocket State Approximation}
The approximation consists in truncating the Hilbert space of
$\:N$--electron wave functions to a finite set of `pocket' basis
states $\:\{\ket{p}\}\:$ ($\:1\le p\le N!\:$). Each pocket state
has one pronounced maximum in configuration space $\:(2\pi)^N\:$.
The locations of the maxima of
$\:|\braket{\vartheta_1,\ldots,\vartheta_N}{p}|\:$ and
$\:|\braket{\vartheta_1,\ldots,\vartheta_N}{p'}|\:$ differ in a
permutation of their coordinates. The eigenstates of
(\ref{1dmodel}) with lowest energies are approximated by linear
combinations of the $\:\{\ket{p}\}\:$, according to the
eigenvectors of the matrix $\:H_{pp'}\equiv\bra{p}H\ket{p'}\:$.
The transformation behaviour under permutations of the
coordinates $\:\{\vartheta_1,\ldots,\vartheta_N\}\:$ fixes
uniquely the total spin $\:S\:$ of the respective Fermion
eigenstate. The off--diagonal elements of $\:H_{pp'}\:$ describe
the exchange processes of electron positions.

The set of symmetrized linear combinations $\:\O_{\Gamma}\{\ket{p}\}\:$
makes the Hamiltonian matrix block diagonal. Here $\:\O_{\Gamma}\:$
is a projector onto states that transform according to the irreducible
representation $\:\Gamma\:$ of the group of permutations of $\:N\:$
elements. Only those blocks are needed and must be diagonalized
which belong to Fermion states of a total electron spin $\:S\:$.
This leads to
\[
n_{\indx S}=\frac{(2S+1)N!}{(N/2+S+1)!(N/2-S)!}
\]
lowest energy levels $\:E_{\indx S}(\phi)\:$ to given spin
$\:S=\{\;\Kg{\mbox{0}\atop\mbox{1/2}},\ldots,N/2\}\:$ \cite{pocket}.

PSA is valid if $\:\Delta\:$ is small compared to the energies
associated with other processes, for instance phonon--like
excitations in the Wigner crystal. The electron `molecule' in a
quantum dot shows exponentially decreasing $\:\Delta\:$ with
increasing $\:a\:$ while the vibrational energies decrease only
according to a power law \cite{hausler91}. PSA is justified at
low densities to determine low energy excitations.

In the present problem rotational excitations have to be
considered additionally, they determine persistent currents.
Without disorder, they scale with the rotational constant
$\:\hbar^2/(2Nm(L/2\pi)^2)=B/N\:$ of $\:N\:$ electron masses.
Rotations by $\:\Delta\vartheta=2\pi a/L\:$ correspond to cyclic
permutations of the electrons and can be incorporated into the
pocket state calculation. Its validity requires that the lowest
energies of long--wavelength phonons should be larger than
rotational excitations
\begin{equation}\label{requ}
\hbar\Omega\equiv 2\pi\frac{e^2}{\ve a_{\rm B}}\frac{a_{\rm B}}{L}
\sqrt{\frac{a_{\rm B}}{a}}
\gg\frac{B}{N}\quad\Longleftrightarrow\quad
s\gg\frac{h}{NmL}\quad\Longleftrightarrow\quad
\sqrt{\frac{a}{a_{\rm B}}}N^2\gg\pi
\end{equation}
where $\:s\:$ is the sound velocity \cite{glazman92a} in the
electron molecule. Condition (\ref{requ}) is on the safe side
to estimate the applicability of the PSA because the impurity
barrier leaves $\:\hbar\Omega\:$ almost unchanged while it
reduces $\:\Delta\:$.
\section{Exchange processes}
In the 1D square well potential the most relevant off--diagonal
entries $\:H_{pp'}\:$ describe the exchange of adjacent electrons
\cite{pocket}. Other matrix elements are exponentially small.
In the ring (\ref{1dmodel}) the importance of the process of circulating
electrons depends on their number and on the strength of the
impurity barrier. I assume here three types of off--diagonal
entries $\:H_{pp'}\:$, $\:p\ne p'\:$. They are illustrated
in Figure~\ref{ring}~:
\begin{itemize}
\item[{\em i.}] The matrix element $\:t\:$ describes the pair
exchange of adjacent electrons on the ring. It leads to the
splitting of the lowest vibrational multiplet into levels of different
total spins. The stronger the electron--electron interaction is, the
more it is difficult for the electrons to pass one another and
the smaller is $\:|t|\:$. It depends also on the width of the ring
(cf.~(\ref{coulomb})) and on the electron density. All of the
$\:t$'s are assumed to be equal, except of the following.
\item[{\em ii.}] The exchange of two electrons located on either
side of the (repulsive) impurity is described by $\:u\:$. In addition
to the Coulomb repulsion the particles have to tunnel through a
barrier, therefore $\:|u|<|t|\:$. Absence of the impurity corresponds
to $\:u=t\:$ and a large impurity makes $\:u\:$ vanishing. Neither
$\:u\:$ nor $\:t\:$ depend on the flux.
\item[{\em iii.}] The (collective) ring exchange transferring all
electrons cyclically by $\:\Delta\vartheta_j=\pm 2\pi a/L\:$ is
described by $\:r\propto {\rm e}^{\pm 2\pi i\phi}\:$. It contains
the same phase factor that the one electron wave function
acquires by turning its coordinate $\:\vartheta_j\longrightarrow
\vartheta_j+2\pi\:$ once around the loop.

An upper limit for $\:|r|\:$ can be deduced from the requirement
that $\:r\:$ should not cause splittings of energies larger than the
rotational constant for a circulating mass $\:Nm\:$
\begin{equation}\label{free}
|r|<\Kg{\d 2\atop 1}\frac{N}{8\pi^2}B\quad\mbox{for}\quad N\;\;
\cases{\!\!\!\!\!&\raisebox{-1.3ex}{even}\cr \!\!\!\!\!&\raisebox{0.7ex}
{odd}\cr}\quad.
\end{equation}
Apart from corrections of order $\:\O(1/N)\:$ (\ref{free}) can be
expressed in terms of the Fermi velocity $\:v_{\rm F}\:$
\begin{equation}\label{fermi}
|r|<\Kg{\d 2\atop 1}\frac{\hbar}{2\pi}\frac{v_{\rm F}}{L}\quad.
\end{equation}
A finite impurity barrier has to be passed by one of the electrons
during the ring exchange. This reduces $\:|r|\:$ compared to (\ref{free})
or (\ref{fermi}). The persistent current, given as a derivative with
respect to $\:\phi\:$, is mainly determined by $\:r\:$.
\end{itemize}
The dependencies and the ranges of $\:t\:$, $\:u\:$ and $\:r\:$
are summarized in Table~\ref{param}. Being tunneling integrals
they are negative ($\:r<0\:$ for $\:\phi=0\:$). Their number
reflects the three relevant parameters in this problem~: The
strengths of the interaction and the impurity, and the
circumference of the ring.

The transport and pinning properties of a one--dimensional Wigner
crystal in the presence of an impurity barrier have been studied
in detail \cite{glazman92a}. The obtained non--trivial
renormalizations of the barrier at zero and at finite
temperatures \cite{barrierrenorm} are consequences of phonon like
excitations in the Wigner crystal which for spinless electrons
are the leading contributions and are of low energy in infinite
systems. In the present consideration vibrational excitations are
neglected and thus, for consistency, the influence of the
strength of the $\:e-e\:$ interaction on $\:r\:$ is ignored.
The results are not valid at temperatures as high as the
vibrational energies.
\section{Estimating \lowercase{$\:|t|,|u|,|r|\:$}}
The dependence of the tunneling integrals $\:t\:$ and $\:u\:$
on the electron density $\:a^{-1}\:$ and on the height $\:V_0\:$ of
the impurity barrier, can be estimated within WKB approximation
\cite{pocket} for $\:b\ll 2\pi/N\:$
\numb
|t|&\sim&\d \exp(-c_1\:\sqrt{a/a_{\rm B}})\\[3ex]
|u|&\sim&\d t\;\exp(-c_2\:\sqrt{a/a_{\rm B}}\:V)\quad.
\nume{esttu}
Here $\:V=V_0/(e^2/\ve a)\:$ is the height of the barrier on
the scale of the Coulomb energy, and
$\:c_1=(1/\pi\sqrt{2})\int_0^{2\pi}{\rm d}\vartheta\;
(|(\pi/\vartheta)-1|)^{-1/2}=1+(\pi+\ln(3+\sqrt{8}))/\sqrt{8}=
2.734\:$ and $\:c_2=1+(\pi-\ln(3+\sqrt{8}))/\sqrt{8}=1.487\:$
are constants determined by the Coulomb interaction (\ref{coulomb}).
The reduction of $\:|r|\:$ can be estimated perturbatively for
small $\:V_0\:$
\begin{equation}\label{estr}
|r|\sim N(B/2\pi-V_0)/2\pi\quad.
\end{equation}
Eqn.~(\ref{estr}) shows that $\:|r|\:$ is necessarily
reduced in the presence of disorder. Experimentally, $\:|r|\:$
is mainly related to the amplitude of the current oscillations
with flux and $\:t\:$ mainly to the temperature dependence as
described in the following section.
\section{Results}
The eigenvalues $\:E_{\indx S}(\phi)\:$ that correspond to
the total electron spin $\:S\:$ can be determined analytically
for $\:S=N/2\:$ (spin polarized states) and for $\:N=3\:$.
\begin{equation}\label{smaxres}
E_{{\indx S}=N/2}(\phi)=(-1)^N2|r|\cos 2\pi\phi\;+\;
(N-1)|t|\;+\;|u|
\end{equation}
corresponds to the energies of {\em spinless} electrons. Their
persistent current is given as $\:I=-\partial E_{\indx S}(\phi)/
\partial(\frac{h}{e}\phi)=(-1)^N(2e/\hbar)|r|\sin 2\pi\phi\:$
which is (also in presence of the impurity) periodic in the
flux quantum $\:h/e\:$. The impurity reduces the amplitude
$\:|r|\:$ of the current oscillations (cf.\ (\ref{estr}))
as compared to the value of one electron circulating with
the Fermi velocity, cf.\ (\ref{fermi}).

The eigenvalues for $\:N=3\:$, $\:S=1/2\:$ are
\numb
\d E_{{\indx S}=1/2}^{(1)}(\phi)&=&\d |r|\cos 2\pi\phi\;-\;
\sqrt{3(|r|\sin 2\pi\phi)^2+(t-u)^2}\\[3ex]
\d E_{{\indx S}=1/2}^{(2)}(\phi)&=&\d |r|\cos 2\pi\phi\;+\;
\sqrt{3(|r|\sin 2\pi\phi)^2+(t-u)^2}\quad.
\nume{n3res}
Many qualitative features can be seen already from the results
(\ref{smaxres}) and (\ref{n3res}). The difference $\:(t-u)\:$
leads to the repulsion between levels of {\em same} spin and
$\:t\:$ separates levels of different spins in energy. For
large $\:|t|\:$ (weak interaction) the ground state energy is
$\:E_{{\indx S}=1/2}^{(1)}(\phi)\:$. The response to small fluxes,
$\:(\partial^2E_{{\indx S}=1/2}^{(1)}(\phi)/\partial\phi^2)|_{\phi=0}=
-|r|((2\pi)^2+12/|t-u|)<0\:$ is paramagnetic, as for three
non--interacting electrons. If $\:|t|<|r|\:$ (strong
interaction) the ground state becomes spin polarized and shows
now diamagnetic response to small fluxes $\:(\partial^2E_{{\indx
S}=3/2}^{(1)}(\phi)/\partial\phi^2)|_{\phi=0}= 8\pi^2|r|>0\:$. In
presence of given impurity $\:|r|\:$ remains constant when the
interaction is increased while $\:|u|\:$ and $\:|t|\:$ ($\:>|u|\:$)
both are reduced (cf.\ Table \ref{param} and Eqn.~(\ref{esttu}))
and the steepness $\:-\partial E_{\indx S}(\phi)/\partial\phi\:$ raises.

In the absence of disorder ($\:t=u\:$) the interaction influences
the magnitude and even the sign of $\:I\:$ if $\:|t|<|r|\:$. The
ground state does not follow a whole segment of the parabola
$\:E_{\indx S}(\phi)\:$ but switches with changing flux to
adjacent pieces of parabolas belonging to other spins.
Only the persistent current of {\em spinless} electrons (within
the approximation considered here) is unaffected by the
electron--electron interaction \cite{weidenmueller93}. Increasing
$\:t=u\:$ increases the distances between the levels, leaving
their $\:\phi$--dispersion almost unchanged. This, eventually,
makes $\:I\:$ independent of a weak interaction $\:|t|\gg |r|\:$.

The eigenvalues for $\:N\ge 4\:$ electrons (Figs.~\ref{n4} and
\ref{n5}) obtained by numerical diagonalizations of Hamiltonian
matrices in the pocket state basis confirm these features.
Figures~~{\sf a} \ refer to `typical' situations where neither
interaction nor disorder dominates. In the Figures~~{\sf b} \ the
electron--electron interaction is increased compared to \ {\sf
a~}, leaving the impurity unchanged. Figures~~{\sf c} \ show
the energy levels in presence of a high impurity barrier but weak
interaction.

Weak interactions lead to pronounced contributions of higher
harmonics to the flux periodicity, see Figures~~{\sf a} \ and
\ {\sf b~}, due to the level repulsion (cf.\ Ref.~\cite{altshuler91}).
A very strong interaction $\:|t|\ll |r|\:$,
however, brings the spin polarized state $\:S=N/2\:$ at
$\:\phi=(1-(-1)^N)/4\:$ close to the ground state.
The $\:h/e$--flux--periodicity (\ref{smaxres}) of the former
is not affected by the impurities. This leads to a non--vanishing
$\:h/e\:$ contribution at strong interaction even after impurity
averaging, in contrast to the purely $\:h/2e$--periodic current
of an odd number of non--interacting electrons.

The low energy of the spin polarized state causes a diamagnetic
response to small fluxes (cf.~(\ref{smaxres})) for $\:|t|\ll |r|\:$
and $\:N\:$ odd ($\:\phi\approx 0\:$). This behaviour is in
contrast to the para\-magnetic susceptibility of an odd number of
non--interacting electrons. But also an even number of electrons
respond diamagnetically if interacting strongly, because the
persistent current approaches $\:h/Ne$--periodicity if $\:t=u\to 0\:$,
as it is found for the 1D--Hubbard model for $\:U\to\infty\:$
\cite{kusmartsev91}. The energy minimum at $\:\phi\approx
N/2N=1/2\:$ (\ref{smaxres}) induces an equivalent minimum
around $\:\phi\approx 0\:$.

In general, the sign of the magnetic susceptibility depends not
only on particle number but also on the disorder and the strength of
the interaction. The long time variations observed in the experiment
\cite{mailly93} can be explained by fluctuating electron numbers but
also by changes in the impurity configuration. In the limit of
both weak disorder and interaction $\:|t|\gapprox |u|\gg
|r|\:$, the sign of the response becomes equal to that of
ballistic electrons on a 1D ring which is diamagnetic only if
$\:N-2\:$ divided by $\:4\:$ is an integer (including
$\:N=2\:$), otherwise paramagnetic \cite{loss91}.

The temperature dependence of the persistent current differs
considerably from non--interacting or spinless electrons. $\:I\:$
vanishes if the temperature exceeds the width of the level
multiplet, which is of the order of $\:\max\{|r|,N|t|\}\:$,
because the trace of the matrix $\:H_{pp'}\:$ is
independent of $\:\phi\:$. The persistent current of interacting
electrons with spin vanishes therefore on an energy scale
$\:\Delta\:$ that varies exponentially with the electron density,
cf.~(\ref{esttu}), while both, the energy $\:\Omega\:$ related
to the sound velocity in the Wigner crystal, and the Fermi
velocity vary only like power laws with $\:a^{-1}\:$.
The former would be relevant to spinless, the latter to
non--interacting electrons.

The impurity cannot smoothen out the discontinuities of the
persistent current at zero temperature at certain values of
$\:\phi\:$, in contrast to spinless electrons. Levels with
different spins can intersect and the magnitude of the current
jumps, mostly also its sign. Simultaneously, the values of the
ground state spins alternate. The experimental observation of
these oscillations would be interesting. Only for very weak
electron--electron interaction $\:|t|\gg |r|\:$ the ground state
spin remains constant, $\:S=0\:$ or $\:S=1/2\:$, Figures~{\sf c}.

Disorder reduces the current for two reasons. At first, the
magnitude $\:|r|\:$ is reduced directly according to
(\ref{estr}). The increasing repulsion between levels of same
spin causes additionally flattened dispersions in $\:\phi\:$ which
reduces the current. An increasing interaction does not influence
the former but reduces significantly the level repulsion and thus
the second reason for current suppression. This can be seen by
comparing the Figures~~{\sf a} \ and \ {\sf b~}. The level
repulsion is reduced because $\:|t|\to 0\:$ forces also $\:u\:$
to vanish. In qualitative agreement with
\cite{weidenmueller93,weidenmueller94} I claim that in the
presence of an impurity the persistent current increases with
increasing electron--electron interaction, though not up to the
value expected for the clean ring. This increase requires,
however, the electron spin.
\section{Summary and Conclusion}
The eigenenergies of a continuous, 1D ring (\ref{1dmodel}) that contains
few, strongly interacting electrons have been analyzed in the presence
of impurity barrier and Aharonov--Bohm flux $\:\phi\;\frac{h}{e}\:$
considering explicitly the electron spin. Correlated,
localized many--electron states have been used to determine
the low energy excitations. The ring is assumed to be
sufficiently so that phonons of the Wigner electron
`molecule' can be ignored. Circular ring symmetry is only used for
some of the estimates, the results do not depend on this assumption.

Spin carring electrons differ qualitatively from spinless electrons
in their persistent current $\:I\:$ (cf.~(\ref{smaxres})).
Impurities do not smoothen the current at zero temperature
if the electron--electron interaction is strong. The spin can
enhance the positive influence of the interaction on the current
interaction can increase $\:I\:$ since no repulsions between levels
of different spins appear --- the spectrum becomes less rigid.
This effect must be distinguished from the current suppression
due to the reduced transmittivity of the impurity barrier which
is not neutralized by strong $\:e-e$--interactions. Therefore
the current does not reach the value $\:I\sim 2ev_{\rm F}/L\:$
found in the absence of both interaction and disorder. The current
remains periodic in once the flux quantum for any $\:N\:$.
Very strong interaction makes the sign of the response to small
fluxes always diamagnetic. This can help to explain the unexpected
recent experimental finding \cite{mailly95} of the diamagnetic
response of an ensemble of semiconducting rings. The temperature
scale on which the persistent current is destroyed depends
exponentially on the electron density. This again is in contrast
to the case of non--interacting or spinless electrons.

The $\:h/Ne$--periodicity of the persistent current found in the
limit of infinitely strong interaction is similar to that of a 1D
Hubbard ring \cite{kusmartsev91,fowler92a}. However, the
suppression of the current, obtained in the lattice model with
increasing interaction between spinless electrons
\cite{montambaux93c}, is not always confirmed in the continuous
model that includes the spin.

It should be noted that magnetic impurities change the
qualitative results presented here only if the rate for spin flip
transitions happen to be comparable to the level repulsions caused
by the fluctuations of the impurity potential so that the spin
states become highly mixed. A weak spin flip scattering is even
implicitly assumed to ensure the system to remain in its ground
state while the applied flux is (slowly) swept through.

Systematic experimental studies of the dependence of the persistent
current on {\em i)} the electron density, {\em ii)} the
height of a hindering tunneling barrier on the ring and {\em iii)}
the temperature would be extremely desirable. Furthermore, it
would be pleasing to observe the electronic ground state spins
to oscillate with flux. The most promising experimental set up
could be the semiconducting ring that allows to regulate the
barrier by a gate, as it has been used in \cite{mailly93}.

\vspace{7mm}\noindent
\begin{minipage}{\textwidth}
{\bf Acknowledgement}\\
The fruitful and stimulating discussions with D.~Loss,
\mbox{H.~Weidenm"uller}, J.~Jefferson, K.~Jauregui and D.~Weinmann
are gratefully acknowledged.
I would like to thank B.~Kramer for numerous valuable impulses and for
continuous encouragement.
Support has been received by the European Community within the
SCIENCE program, grants SCC$^*$--CT90--0020 and CHRX--CT93--0126.
\end{minipage}
\newpage
\newcommand{\pap}[5]{{\sc #1}, \ #2 {\bf #3} (19#5) #4}
\newcommand{\bk}[4]{{\sc #1} in {\it #2\/}, #3, 19#4}
\renewcommand{\baselinestretch}{0.9}\large
\newpage
\begin{table}[htb]
\begin{tabular}{|@{\hspace*{2mm}}c@{\hspace*{2mm}}|
*{3}{@{\hspace*{2mm}}c@{\hspace*{2mm}}|}}
\hline
\rule{0ex}{5ex}{\large\sf parameter} & {\large\sf depends on the} &
{\large\sf is small for} & {\large\sf maximum value}\\[3ex] \hline
\rule{0ex}{5ex}{\large $|r|$} & {\large\sf impurity} & {\large\sf
strong impurity} & {\large $NB/(2\pi)^2$}\\[3ex] \hline
\rule{0ex}{5ex}{\large $t$} & {\large\sf interaction} & {\large\sf strong
interaction} & {\large $\:|t|\gg |r|\:$}\\[3ex] \hline
\rule{0ex}{5ex}{\large $u$} & {\large\sf impurity} & {\large\sf strong
impurity} & {\large $u=t$}\\[3ex] \hline
\end{tabular}\vspace{1cm}
\caption[param]{\label{param}\small
Magnitudes of the dominant tunneling integrals within the pocket
state description.
}
\end{table}
\newpage
\noindent\begin{figure}[h]
\parbox[b]{14cm}
{\caption[ring]{\label{ring}
Illustration of the exchange processes associated with the
tunneling integrals $\:r\:$, $\:t\:$ and $\:u\:$ (see text).
}}
\end{figure}
\noindent\begin{figure}[h]
\parbox[b]{14cm}
{\caption[n4]{\label{n4}
Energy levels versus the magnetic flux $\:\phi\;\frac{h}{e}\:$ for
$\:N=4\:$ electrons. Thick solid lines~: $\:S=2\:$, dotted lines~:
$\:S=1\:$, thin solid lines~: $\:S=0\:$.

Parameters are \
{\sf a~:} $\:|r|=1\:$,$\:t=-3/4\:$,$\:u=-1/12\:$;
\ {\sf b~:} $\:|r|=1\:$,$\:t=-1/4\:$,$\:u=-1/12\:$;
\ {\sf c~:} $\:|r|=1/3\:$,$\:t=-1\:$,$\:u=-1/3$ (see text and
Figure~\ref{ring}). Figures \ {\sf b~} correspond to strong
interaction and Figures \ {\sf c~} to strong impurity.

Below each Figure $\:-\partial E_0(\phi)/\partial\phi\:$ of
the ground state energy $\:E_0\:$ is plotted
which is proportional to the persistent current at zero
temperature.
}}
\end{figure}
\noindent\begin{figure}[h]
\parbox[b]{14cm}
{\caption[n5]{\label{n5}
Same as Figures~\ref{n4} for the same parameters
but for $\:N=5\:$ electrons. Thick solid lines~:
$\:S=5/2\:$, dotted lines~: $\:S=3/2\:$, thin solid
lines~: $\:S=1/2\:$.
}}
\end{figure}

\normalsize
\begin{thebibliography}{99}
\bibitem{percurr}\pap{F.~Hund}{Annalen der Physik (Leipzig)}{32}{102}{38},\\
 \pap{F.~Bloch}{Phys.~Rev.}{166}{415}{68},
 \pap{M.~B"uttiker, Y.~Imry, R.~Landauer}{Phys.~Lett}{96 A}{365}{83}.
\bibitem{mailly93}\pap{D.~Mailly, C.~Chapelier, A.~Benoit}
 {Phys.~Rev.~Lett.}{70}{2020}{93}.
\bibitem{levi}\pap{L.~P.~L\'evy, G.~Dolan, J.~Dunsmuir, H.~Bouchiat}
 {Phys.~Rev.~Lett.}{64}{2074}{90}.
\bibitem{chandrasekar}\pap{V.~Chandrasekhar, R.~A.~Webb, M.~J.~Brady,
 M.~B.~Ketchen, W.~J.~Gallaghar, A.~Kleinsasser}{Phys.~Rev.~Lett.}
 {67}{3578}{91}.
\bibitem{oppen91}\pap{F.~von Oppen, E.~K.~Riedel}
 {Phys.~Rev.~Lett.}{66}{84}{91}.
\bibitem{altshuler91}\pap{B.~L.~Altshuler, Y.~Gefen, Y.~Imry}
 {Phys.~Rev.~Lett.}{66}{88}{91}.
\bibitem{eckern93}\pap{V.~Ambegaokar, U.~Eckern}{Phys.~Rev.~Lett.}
 {65}{381}{90},\\
 \pap{U.~Eckern, A.~Schmid}{Annalen der Physik}{2}{180}{93} and references
 therein.
\bibitem{ando82}
 \pap{T.~Ando, A.~B.~Fowler, F.~Stern}{Rev.~Mod.~Phys.}{54}{437}{82}.
\bibitem{hausler93}\pap{W.~H"ausler, B.~Kramer}{Phys.~Rev.~B}{47}{16353}{93}.
\bibitem{loss92}\pap{D.~Loss}{Phys.~Rev.~Lett.}{69}{343}{92},\\
 \bk{D.~Loss, D.~L.~Maslov}{Quantum Dynamics of
 Submicron Structures}{ed.~by H.~A.~Cerdeira, B.~Kramer, G.~Sch"on,
 NATO ASI Series, Vol.~291, Kluwer, Dordrecht}{95};\\
 cf.\ also the review by \
 \pap{A.~A.~Zvyagin, I.~V.~Krive}{Low.~Temp.~Phys.}{21}{533}{95}.
\bibitem{weidenmueller93}\pap{A.~M"uller-Groeling, H.~A.~Weidenm"uller,
 C.~H.~Lewenkopf}{Europhys.~Lett.}{22}{193}{93},\\
 \pap{H.~A.~Weidenm"uller}{Physica}{A~200}{104}{93}.
\bibitem{montambaux93c}
 \pap{G.~Bouzerar, D.~Poilblanc, G.~Montambaux}{Phys.~Rev.~B}{49}{8258}{94}.
\bibitem{fowler92a}\pap{N.~Yu, M.~Fowler}{Phys.~Rev.~B}{45}{11795}{92}.
\bibitem{berkovits94}
 \pap{R.~Berkovits}{Europhys.~Lett.}{25}{681}{94},\\
 \pap{D.~Poilblanc, T.~Ziman, J.~Bellissard,
 F.~Mila, G.~Montambaux}{Europhys.~Lett.}{22}{537}{93},\\
 \pap{M.~Faas, B.~D.~Simons, X.~Zotos, B.~L.~Altshuler}
 {Phys.~Rev.~B}{48}{5439}{93}.
\bibitem{imry91}{\sc Y.~Imry} in {\em Quantum Coherence in Mesoscopic
 Systems}, ed.\ by B.~Kramer, NATO ASI Series B {\bf 254}, Plenum
 Press, New York (1991).
\bibitem{pocket} {\sc W.~H"ausler} in {\it Festk"orperprobleme~:
 Advances in Solid state physics}, volume 34, Vieweg Verlag,
 Braunschweig (1994) and accepted for publication in Z.~Phys.~B.
\bibitem{hausler91}
 \pap{W.~H"ausler, B.~Kramer, J.~Ma\v{s}ek}{Z.~Phys.~B}{85}{435}{91}.
\bibitem{glazman92a}\pap{L.~I.~Glazman, I.~M.~Ruzin, B.~I.~Shklovskii}
 {Phys.~Rev.~B}{45}{8454}{92}.
\bibitem{barrierrenorm}\pap{A.~I.~Larkin and P.~A.~Lee}{Phys.~Rev.~B}
 {17}{1596}{78},\\
 \pap{A.~O.~Gogolin, N.~V.~Prokof'ev}{Phys.~Rev.~B}{50}{4921}{94},\\
 \pap{I.~V.~Krive, P.~Sandstr\"om, R.~I.~Shekhter, S.~M.~Girvin, M.~Jonson}
 {Phys.~Rev.~B}{52}{16451}{95}.
\bibitem{kusmartsev91}
 \pap{F.~V.~Kusmartsev}{J.~Phys.: Condens.\ Matter}{3}{3199}{91}.
\bibitem{loss91}\pap{D.~Loss, P.~Goldbart}{Phys.~Rev.~B}{43}{13762}{91}.
\bibitem{weidenmueller94}
 \pap{A.~M"uller-Groeling, H.~A.~Weidenm"uller}{Phys.~Rev.~B}{49}{4752}{94}.
\bibitem{mailly95}\pap{B.~Reulet, M.~Ramin, H.~Bouchiat, and D.~Mailly}
 {Phys.~Rev.~Lett.}{75}{124}{95}.
\end{thebibliography}
\end{document}